\documentclass[10pt,english,natbib]{paper}
\usepackage[T1]{fontenc}
\usepackage[latin9]{inputenc}
\usepackage{prettyref}
\usepackage{amsmath}
\usepackage{amssymb}

\makeatletter

\providecommand{\tabularnewline}{\\}

\makeatother

\usepackage{babel}
\begin{document}

\title{Formalization of some central theorems in combinatorics of finite
sets}

\author{Abhishek Kr Singh\\Tata Institute of Fundamental Research, Mumbai }

\maketitle
\begin{abstract}
We present fully formalized proofs of some central theorems from combinatorics.
These are Dilworth's decomposition theorem, Mirsky's theorem, Hall's
marriage theorem and the Erd\H{o}s-Szekeres theorem. Dilworth's decomposition
theorem is the key result among these. It states that in any finite
partially ordered set (poset), the size of a smallest chain cover
and a largest antichain are the same. Mirsky's theorem is a dual of
Dilworth's decomposition theorem, which states that in any finite
poset, the size of a smallest antichain cover and a largest chain
are the same. We use Dilworth's theorem in the proofs of Hall's Marriage
theorem and the Erd\H{o}s-Szekeres theorem. The combinatorial objects
involved in these theorems are sets and sequences. All the proofs
are formalized in the Coq proof assistant. We develop a library of
definitions and facts that can be used as a framework for formalizing
other theorems on finite posets. 
\end{abstract}

\section{Introduction}

Formalization of any mathematical theory is a difficult task because
the length of a formal proof blows up significantly. In combinatorics
the task becomes even more difficult due to the lack of structure
in the theory. Some statements often admit more than one proof using
completely different ideas. Thus, exploring dependencies among important
results may help in identifying an effective order amongst them. Dilworth's
decomposition theorem, first proved by R.~P.~ Dilworth\cite{key-1}
in 1951, is a well-known result in combinatorics. It states that in
any finite partially ordered set (poset) the size of a smallest chain
cover and a largest antichain are the same. Since then, the theorem
attracted significant attention and several new proofs \cite{key-2,key-3,key-4}
were discovered. In addition to being an important structural result
on posets, Dilworth's Theorem can be used to give intuitive and concise
proofs of some other important results in combinatorics such as Hall's
Theorem \cite{key-6,key-7}, the Erd\H{o}s-Szekeres Theorem \cite{key-8},
and Konig's Theorem \cite{key-19}. 

In this paper we present a fully formalized proof of Dilworth's decomposition
theorem. Among the several proofs available we follow the proof by
Perles \cite{key-2} due to its clean and concise reasoning steps.
We then mechanize proofs of Hall's Marriage theorem \cite{key-6,key-7}
and the Erd\H{o}s-Szekeres theorem \cite{key-8}. Proofs that we mechanize
for these theorems essentially use Dilworth's decomposition theorem.
In these proofs a finite poset is constructed from the objects involved
and then Dilworth's decomposition theorem is applied to obtain the
result. These proofs are explained in detail in Section 3-4. We also
formalize a dual of Dilworth's Theorem (Mirsky's Theorem \cite{key-17})
which relates the size of an antichain cover and a chain in a poset. 

Formalization of known mathematical results can be traced back to
the systems Automath and Mizar \cite{key-13}. Mizar hosts the largest
repository of formalized mathematics. Mizar system also supports some
built in automation to save time during proof development. However,
this results in a large kernel (core) and reduces our faith in the
system. The Coq proof assistant deals with this problem in a novel
way. It separates the process of proof development from proof checking.
Some small scale proof automation is also possible in Coq. However,
every proof process finally yields a proof-term which is verified
using a small kernel. Thus the part (kernel) of the code we need to
trust remains small. All the results discussed in this paper are fully
formalized in the Coq proof assistant. In addition to a small kernel,
the Coq proof assistant also has some other useful features such as
\emph{dependent records} and \emph{coercions}. Dependent records are
used to pack mathematical objects and their properties in one definition.
For example, in the Coq standard library different components of a
partial order and their properties are expressed using a single definition
of dependent record (PO). Similarly, coercions can be used to define
a hierarchy among mathematical structures. This avoids redefining
similar things at different places. The Coq system also hosts a standard
library \cite{key-14} that contains a large collection of useful
definitions and results. We use this facility and avoid new definitions
unless absolutely essential. 

In this paper, we present the details of our mechanized proofs of
Dilworth's, Mirsky's, Hall's, and the Erd\H{o}s-Szekeres theorems.
All the terms that appear in the formal statement of these theorems
are explained in Section~2 and Sections~4-6. The exact definitions
of these terms in Coq are listed in \prettyref{sec:Appendix} (Appendix).
Description of some useful results on sets and posets appears in \prettyref{sec:Some-useful-results}.
Finally, we review related work in \prettyref{sec:Related-Work} and
conclude in \prettyref{sec:Conclusions}. 

\section{Definitions }

Once a statement is proved in Coq, the proof is certified without
having to go through the proof-script. It is however necessary to
verify whether the statement being proved correctly represents the
original theorem. Therefore the number of new definitions needed to
understand the theorem statement should be small. We have attempted
to achieve this by reusing the definitions from the Coq standard Library
whenever possible. In this section we explain the definitions of all
the terms which appear in the formal statements of Dilworth's and
Mirsky's Theorem. 

\subsection{Definitions from the Standard Library}

The Coq Standard Library\cite{key-14} is well documented. We have
used the \emph{Sets }module from the Standard Library, where a declaration
S: Ensemble U is used to represent a set $S$. 
\begin{itemize}
\item Sets are treated as predicates, i.e, $x\in S$ iff S x is provable. 
\item Set membership is written as In S x instead of just writing S x. 
\item The Empty set is defined as a predicate Empty\_set which is not provable
anywhere. Singleton x and Couple x y represent the sets $\{x\}$ and
\{$x,y\}$ respectively. 
\end{itemize}
A Partial Order is defined as a record type in the Coq standard library.
It has four fields,\texttt{ }
\begin{description}
\item [{Record}] PO (U : Type) : Type := Definition\_of\_PO \{ \\
Carrier\_of : Ensemble U; \\
Rel\_of : Relation U;\\
 PO\_cond1 : Inhabited U Carrier\_of;\\
 PO\_cond2 : Order U Rel\_of \}.
\end{description}
For example, consider the following declaration,
\begin{description}
\item [{Variable}] U:Type.
\item [{Variable}] P: PO U. 
\end{description}
It creates a record P of type PO U. Here P can be treated as a poset
with four fields. The first field of P is accessed using the term
Carrier\_of \_ P. It represents the carrier set of P. The second field
represents binary relation $\leq$ of the partially ordered set P.
It is accessed using the term Rel\_of \_ P. The term PO\_cond1 \_
P is a proof that the carrier set of P is a non-empty set. Similarly,
the term PO\_cond2 \_ P is a proof that $\leq$ is an order (i.e,
reflexive, transitive and antisymmetric). 

\subsection{New Definitions}

\subsubsection*{Coercions and Finite partial orders}

We extend the definition of poset to define \emph{finite partial orders}
(FPO) as a dependent record,
\begin{description}
\item [{Record}] FPO (U : Type) : Type := Definition\_of\_FPO \{\\
PO\_of :> PO U ; \\
FPO\_cond : Finite \_ (Carrier\_of \_ PO\_of ) \}.
\end{description}
It has two components; a partial order and a proof that the carrier
set of the partial order is finite. Here, FPO is defined as a dependent
record which inherits all the fields of type PO. Note the use of coercion
symbol \texttt{:>} in defining the first field of the record FPO.
Here, PO\_of acts as a function and is applied automatically to any
term of type FPO that appears in a context where a term of type PO
is expected. Hence, from now onward we can use an object of type FPO
in any context where an object of type PO is expected. 

\subsubsection*{Chains and antichains as predicates}

In the Coq Standard Library a chain is defined as a poset whose carrier
set is totally ordered.
\begin{description}
\item [{Record}] Chain : Type := Definition\_of\_chain \{\\
PO\_of\_chain : PO U; \\
Chain\_cond : Totally\_ordered U PO\_of\_chain (@Carrier\_of \_ PO\_of\_chain)\}.
\end{description}
However, using this definition it becomes difficult to say that a
given set is a chain in two different posets. In the proof of Dilworth's
theorem we frequently refer to a set in the context of two different
posets and wish to claim that the set is totally ordered in both the
posets. Thus we use a different definition for chain. A chain is defined
using a predicate Is\_a\_chain\_in. For a finite partial order P:
FPO U\texttt{ }on some type \texttt{U} let, C := Carrier\_of U P and\texttt{
}R:= Rel\_of U P. Then, 
\begin{description}
\item [{Definition}] Is\_a\_chain\_in (e: Ensemble U): Prop:= (Included
U e C /\textbackslash{} Inhabited U e) /\textbackslash{} ($\forall$
x y:U, (Included U (Couple U x y) e) $\rightarrow$ R x y \textbackslash{}/
R y x). 
\end{description}
Here, a \emph{chain} is a subset of P any two of whose elements are
comparable. A subset of P in which no two distinct elements are comparable
is called an \emph{antichain}. An antichain is defined using the predicate
Is\_an\_antichain\_in. Note that, a chain and an antichain can have
at most one element in common. In a similar way we also define the
following notions,
\begin{itemize}
\item A\emph{ chain cover} is a collection of chains whose union is the
entire poset. 
\item An \emph{antichain cover }is a collection of antichains such that
their union is the entire poset. 
\item The \emph{width} of a poset P, $width(P)$, is the size of a largest
antichain in P. 
\item The \emph{height} of a poset, $height(P),$ is the size of a largest
chain in P. 
\item An element $b\in P$ is called a \emph{maximal element }if there is
no $a\in P$ such that $b\leq a$. 
\item An element $a\in P$ is called a \emph{minimal element} if there is
no $b\in P$ such that $b\leq a$. 
\end{itemize}
The exact definitions that we use for these terms are listed in \prettyref{sec:Appendix}(Appendix). 

\section{Some useful results on sets and posets \label{sec:Some-useful-results}}

In this section we explain some general results on finite partial
orders. These results are used at more than one place in the formal
proofs of these theorems. They are proved as Lemmas and compiled in
separate files. Most of the Lemma's statements can be inferred from
their names. These Lemmas appear with the same name in the actual
Coq files. Here we only provide an English description of some of
them. 

\subsection*{Existence proofs}

A large number of lemmas are concerned with the existence of a defined
object. For example, in our proof when we say ``Let A be an antichain
of the poset P...'' we assume that there exists an antichain for
the poset P. However, in a formal system like Coq, we need a proof
of existence of such an object before we can instantiate it. Following
is a partial list of such results: 
\begin{description}
\item [{Lemma-1}] \emph{Chain\_exists}: There exists a chain in every finite
partial order (FPO). \textbf{}\\
\textbf{Proof.} Trivial.
\item [{Lemma-2}] \emph{Chain\_cover\_exists:} There exists a chain cover
for every FPO.\\
\textbf{Proof.} Trivial.
\item [{Lemma-3}] \emph{Minimal\_element\_exists:} The set minimal(P) is
non-empty for every P: FPO. \textbf{}\\
\textbf{Proof.} Using induction on the size of P.
\item [{Lemma-4}] \emph{Maximal\_element\_exists:} The set maximal(P) is
non-empty for every P: FPO. \textbf{}\\
\textbf{Proof.} Using induction on the size of P.
\item [{Lemma-5}] \emph{Largest\_element\_exists:} If a finite partial
order is also totally ordered then there exists a largest element
in it. \textbf{}\\
\textbf{Proof.}\textbf{\emph{ }}The maximal element becomes the largest
element and we know that there exists a maximal element. 
\item [{Lemma-6}] \emph{Minimal\_for\_every\_y:} For every element $y$
of a finite partial order P there exists an element $x$ in P such
that $x\leq y$ and $x\in\text{minimal(P)}$. \emph{}\\
\textbf{Proof.} Let $X=\{x:P|\,x\le y\}$. Then the poset $(X,\leq)$
will have a minimal element, say $x_{0}$. It is also a minimal element
of P. 
\item [{Lemma-7}] \emph{Maximal\_for\_every\_x:} For every element $x$
of a finite partial order P there exists an element $y$ in P such
that $x\leq y$ and $y\in\text{maximal(P)}$. \emph{}\\
\textbf{Proof.} Let $Y=\{y:P|\,x\le y\}$. Then the poset $(Y,\leq)$
will have a maximal element, say $y_{m}$. It is also a maximal element
of P. 
\item [{Lemma-8}] \emph{Largest\_set\_exists:} There exists a largest set
(by cardinality) in a finite and non-empty collection of finite sets.\textbf{\emph{
}}\textbf{}\\
\textbf{Proof. }Consider the collection of sets together with the
strict set-inclusion relation. This forms a finite partial order.
Any maximal element of this finite partial order will be a largest
set. Moreover, such a maximal element exists due to Lemma-4.
\item [{Lemma-9}] \emph{exists\_largest\_antichain:} In every finite partial
order there exists a largest antichain.\emph{ }\textbf{}\\
\textbf{Proof.} Note that this statement is not true for partial orders.
The proof is similar to Lemma-8. 
\item [{Lemma-10}] \emph{exists\_largest\_chain:} In every finite partial
order there exists a largest antichain.\emph{ }\textbf{}\\
\textbf{Proof.} Again, it is true only for finite partial orders.
Proof is similar to Lemma-8.
\end{description}

\subsection*{Some other proofs}

When dealing with sets the set-inclusion relation occurs more naturally
than the comparison based on the set sizes. Therefore, we defined
a binary relations \emph{Inside} (or $\prec$) on the collection of
all the finite partial orders. 
\begin{itemize}
\item We say $P_{1}\prec P_{2}$ iff carrier set of $P_{1}$ is strictly
included in the carrier set of $P_{2}$ and both the posets are defined
on the same binary relation. 
\end{itemize}
In order to use well-founded induction we proved that the relation
$\prec$ is well founded. 
\begin{description}
\item [{Lemma-11}] \emph{Inside\_is\_WF:} The binary relation Inside (i.e,
$\prec$ ) is well founded on the set of all finite partial orders.
\textbf{}\\
\textbf{Proof.} Using strong induction on the size of finite partial
orders. 
\item [{Lemma-12}] \emph{Largest\_antichain\_remains:} If $\mathcal{A}$
is a largest antichain of $P_{2}$ and $P_{1}\prec P_{2}$ then $\mathcal{A}$
is also a largest antichain in $P_{1}$ provided $\mathcal{A}\subset P_{1}$.
\textbf{}\\
\textbf{Proof. }Assume otherwise, then there will be a larger antichain
say $\mathcal{A}'$ in $P_{1}$. This will also be larger in $P_{2}$,
which contradicts.
\item [{Lemma-13}] \emph{NoTwoCommon:} A chain and an antichain can have
at most one element in common. \textbf{}\\
\textbf{Proof.} Trivial.
\item [{Lemma-14}] \emph{Minimal\_is\_antichain:} Minimal(P) is an antichain
in P. \textbf{}\\
\textbf{Proof.} Trivial.
\item [{Lemma-15}] \emph{Maximal\_is\_antichain:} Maximal(P) is an antichain
in P. \textbf{}\\
\textbf{Proof. }Trivial. 
\item [{Lemma-17}] \emph{exists\_disjoint\_cover:} If $\mathcal{C_{V}}$
is a smallest chain cover of size $m$ for P, then there also exists
a disjoint chain cover $\mathcal{C_{V}}'$ of size $m$ for P. \textbf{}\\
\textbf{Proof.} Using induction on $m$.
\item [{Lemma-18}] \emph{Largest\_chain\_has\_maximal:} In any finite poset
P, $\text{maximal(P)}$ shares an element with every largest chain
of P. \textbf{}\\
\textbf{Proof. }First we observe that every chain in a finite poset
has a largest element. We prove that this element is also in $\text{maximal(P)}$. 
\item [{Lemma-19}] \emph{Largest\_chain\_has\_minimal:} In any finite poset
P, $\text{minimal(P)}$ shares an element with every largest chain
of P. \textbf{}\\
\textbf{Proof. }Similar to the proof of Lemma-18. 
\item [{Lemma-20}] \emph{Pre\_ES:} If P is a poset with $r.s+1$ elements,
then it has a chain of size $r+1$ or an antichain of size $s+1$.\\
\textbf{Proof.} There can be two cases; either there is an antichain
$\mathcal{A}$ of size $s+1$ or the size of a largest antichain is
$s$. In the first case statement is trivially true. In the second
case, using Dilworth's theorem we know that there exists a chain cover
$\mathcal{C_{V}}$ of size $s$. Since $\mathcal{C_{V}}$ covers the
whole poset P and its size is $r.s+1$, there must be an chain of
size at least $r+1$ in $\mathcal{C_{V}}$. 
\end{description}

\section{Mirsky's theorem and Dilworth's decomposition theorem}

\subsection{Mirsky's theorem}

Mirsky's theorem relates the size of an antichain cover and a chain
in a poset. The definitions we have seen so far are sufficient to
express the formal statement of Mirsky's theorem in Coq.
\begin{description}
\item [{Theorem}] Dual\_Dilworth: $\forall$ (P: FPO U), Dual\_Dilworth\_statement
P.
\end{description}
where, Dual\_Dilworth\_statement is defined as,
\begin{description}
\item [{Definition}] Dual\_Dilworth\_statement:= fun (P: FPO U) $\Rightarrow$
$\forall$ (m n: nat), (Is\_height P m) $\rightarrow$ ($\exists$
cover: Ensemble (Ensemble U), (Is\_a\_smallest\_antichain\_cover P
cover) /\textbackslash{} (cardinal \_ cover n)) $\rightarrow$ m=n.
\end{description}
It states that in any poset the maximum size of a chain is equal to
the minimum number of antichains in any antichain cover. In other
words, if $c(P)$ represents the size of a smallest antichain cover
of P, then $height(P)=c(P)$. \textbf{}\\
\textbf{Proof} : The equality will follow if one can prove:
\begin{enumerate}
\item Size of a chain $\leq$ Size of an antichain cover, and
\item There is an antichain cover of size equal to $height(P)$. 
\end{enumerate}
It is easy to see why (1) is true. Any chain shares at most one element
with each antichain from an antichain cover. Moreover, every element
of the chain must be covered by some antichain from the antichain
cover. Hence, the size of any chain is smaller than or equal to the
size of any antichain cover. 

We will prove (2) using strong induction on the size of the largest
chain of $P$. Let $m$ be the size of the largest chain in P, i.e,
$m=height(P)$. 
\begin{itemize}
\item Induction hypothesis: For all posets $P'$ of height at most $m-1$,
there exists an antichain cover of size equal to $height(P')$.
\end{itemize}
Induction Step: Let $M$ denote the set of all maximal elements of
$P$, i.e, $M=\text{maximal(P)}$. Observe that $M$ is a non-empty
antichain and shares an element with every largest chain of $P$.
Consider now the partially ordered set $(P-M,\leq)$. The length of
the largest chain in $P-M$ is at most $m-1$. On the other hand,
if the length of the largest chain in $P-M$ is less than $m-1$,
$M$ must contain two or more elements that are members of the same
chain, which is a contradiction. Hence, we conclude that the length
of largest chain in $P-M$ is $m-1$. Using induction hypothesis there
we get an antichain cover $\mathcal{A_{C}}$ of size $m-1$ for $P-M$.
Thus, we get an antichain cover $\mathcal{A_{C}}\cup\{M\}$ of size
$m$ for $P$. $\square$ 

Note that in the induction step of the above proof we assume that
$\text{maximal(P)}$ shares an element with every largest chain of
P. However, in the formal setting we need a proof of this fact. It
is proved as Lemma-18 in \prettyref{sec:Some-useful-results}.

\subsection{Dilworth's decomposition theorem}

Dilworth's decomposition theorem is the central result in our formalization.
It relates the size of a chain cover and an antichain in a poset.
We prove the following formal statement,
\begin{description}
\item [{Theorem}] Dilworth: $\forall$ (P: FPO U), Dilworth\_statement
P.
\end{description}
where Dilworth\_statement is defined as, 
\begin{description}
\item [{Definition}] Dilworth\_statement:= fun (P: FPO U)$\Rightarrow$
$\forall$ (m n: nat), (Is\_width P m) $\rightarrow$ ($\exists$
cover: Ensemble (Ensemble U), (Is\_a\_smallest\_chain\_cover P cover)
/\textbackslash{} (cardinal \_ cover n)) $\rightarrow$ m=n.
\end{description}
It states that in any poset, the maximum size of an antichain is equal
to the minimum number of chains in any chain cover. In other words,
if $c(P)$ represents the size of a smallest chain cover of P, then
$width(P)=c(P)$. 

The statement of Dilworth's theorem appears dual to the statement
of Mirsky's theorem. However, the proof of Dilworth's theorem is more
involved. The key idea in proving Mirsky's theorem was to identify
an antichain which intersects every largest chain (Lemma-18). It is
however not easy to identify a chain in a poset which intersects every
largest antichain. This is the main difficulty in translating the
proof of Mirsky's theorem to a proof of Dilworth's theorem. Therefore,
we mechanize a different proof of Dilworth's theorem due to Perles\cite{key-2}.\textbf{}\\
\textbf{Proof} (Perles): The equality $width(P)=c(P)$ will follow
if one can prove:
\begin{enumerate}
\item Size of an antichain $\leq$ Size of a chain cover, and
\item There is a chain cover of size equal to $width(P)$.
\end{enumerate}
Again, it is easy to see why (1) is true. Assume otherwise, i.e.,
there is an antichain $\mathcal{A}$ of size bigger than the size
of a smallest chain cover $\mathcal{C_{V}}$. Then $\mathcal{A}$
will have more elements than the number of chains in $\mathcal{C_{V}}.$
Hence, there must exist a chain $\mathcal{C}$ in $\mathcal{C_{V}}$
which covers two elements of $\mathcal{A}$. However, this cannot
be true since a chain and an antichain (in this case $\mathcal{C}$
and $\mathcal{A}$) can have at most one element in common. 

Proof of (2) is more involved. We will prove (2) using strong induction
on the size of $P$. Let $m$ be the size of the largest antichain
in P, i.e., $m=width(P)$. 
\begin{itemize}
\item Induction hypothesis: For all posets $P'$ of size at most $n$, there
exists a chain cover of size equal to $width(P')$.
\end{itemize}
Induction step: Fix a poset P of size at most $n+1$. Let maximal(P)
and minimal(P) represent respectively the set of all maximal and the
set of all minimal elements of P. Now, one of the following two cases
might occur,
\begin{enumerate}
\item There exists an antichain $\mathcal{A}$ of size $m$ which is neither
maximal(P) nor minimal(P).
\item No antichain other than maximal(P) or minimal(P) has size $m$. 
\end{enumerate}
\textbf{Case-1:} For the first case we define the sets $P^{+}$ and
$P^{-}$ as follows:
\begin{center}
\begin{tabular}{c}
\tabularnewline
$P^{+}=\left\{ x\in P:\,x\geq y\,\,\text{for some }y\in\mathcal{A}\right\} $\tabularnewline
$P^{-}=\left\{ x\in P:\,x\leq y\,\,\text{for some }y\in\mathcal{A}\right\} $\tabularnewline
\tabularnewline
\end{tabular}
\par\end{center}

Here $P^{+}$ captures the notion of being above $\mathcal{A}$ and
$P^{-}$ captures the notion of being below $\mathcal{A}$. Note that
the elements of $\mathcal{A}$ are both above and below $\mathcal{A}$,
i.e, $\mathcal{A}\subseteq P^{+}\cap P^{-}$. For any arbitrary element
$x\in P$
\begin{itemize}
\item If $x\in A$ then $x\in P^{+}\cap P^{-}$ and hence $x\in P^{+}\cup P^{-}$.
\item If $x\notin\mathcal{A}$ then $x$ must be comparable to some element
in $\mathcal{A}$; otherwise $\{x\}\cup\mathcal{A}$ will be an antichain
of size $m+1$. Hence, if $x\notin\mathcal{A}$ then $x\in P^{+}\cup P^{-}$. 
\end{itemize}
Therefore, $P^{+}\cup P^{-}=P$. Since there is at least one minimal
element not in $\mathcal{A}$, $P^{+}\neq P$. Similarly $P^{-}\neq P$.
Thus $|P^{+}|<|P|$ and $|P^{-}|<|P|$, hence we will be able to apply
induction hypothesis to them. Observe that $\mathcal{A}$ is also
a largest antichain in the poset restricted to $P^{+}$; because if
there was a larger one, it would have been larger in $P$ also. Therefore
by induction there exists a chain cover of size $m$ for $P^{+}$,
say $P^{+}=\cup_{i=1}^{m}C_{i}$. Similarly, there is a chain cover
of size $m$ for $P^{-}$, say $P^{-}=\cup_{i=1}^{m}D_{i}$.

Elements of $\mathcal{A}$ are the minimal elements of the chains
$C_{i}$ and the maximal elements of the chains $D_{i}$. Therefore
we can join the chains $C_{i}$ and $D_{i}$ together in pairs to
form $m$ chains which form a chain cover for the original poset $P$.

\textbf{Case-2: }In this case we can't have an antichain of size $m$
which is different from both maximal(P) and minimal(P). Consider a
minimal element $x.$ Choose a maximal element $y$ such that $x\leq y$.
Such a $y$ always exists. Remove the chain $\{x,y\}$ from $P$ to
get the poset $P'$. Then $P'$ contains an antichain of size $m-1$.
Also note that $P'$ can't have an antichain of size $m$. Because
if there was an antichain of size $m$ in $P'$, then that would also
be an antichain in $P$ which is different from both maximal(P) and
minimal(P), and we would have been in the first case (i.e., Case-1).
Hence by induction hypothesis we get a chain decomposition of $P'$
of size $m-1$. These chains, together with $\{x,y\}$, give a decomposition
of $P$ into $m$ chains. $\square$ 

We mechanize the above proof in Coq with a slight modification. Instead
of using induction on the cardinality of posets we use well-founded
induction on the strict set-inclusion relation. When working with
the Ensemble module of the Coq standard library it is easy to deal
with the set-inclusion relation compared to the comparison based on
set cardinalities. Thus, we defined a binary relations \emph{Inside}
(or $\prec$) on the collection of all the finite partial orders. 
\begin{itemize}
\item We say $P_{1}\prec P_{2}$ iff carrier set of $P_{1}$ is strictly
included in the carrier set of $P_{2}$ and both the posets are defined
on the same binary relation. 
\end{itemize}
Then to use well-founded induction we proved that the relation $\prec$
is well founded. This is explained as Lemma-11 in \prettyref{sec:Some-useful-results}. 

In the formalization of above proofs we use the principle of excluded
middle at many places. At certain points, we also need to extract
functions from relations. Therefore, we import the \emph{Classical}
and \emph{ClassicalChoice} modules of the standard library, which
assumes the following three axioms:
\begin{description}
\item [{Axiom}] classic : $\forall$ P:Prop, P \textbackslash{}/ \textasciitilde{}
P.
\item [{Axiom}] dependent\_unique\_choice : $\forall$ (A:Type) (B:A $\rightarrow$
Type) (R:$\forall$ x:A, B x $\rightarrow$ Prop), ($\forall$ x :
A, $\exists$! y : B x, R x y) $\rightarrow$($\exists$ f : ($\forall$
x:A, B x), $\forall$ x:A, R x (f x)).
\item [{Axiom}] relational\_choice : $\forall$ (A B : Type) (R : A$\rightarrow$B$\rightarrow$Prop),
($\forall$ x : A, $\exists$ y : B, R x y) $\rightarrow$ $\exists$
R' : A$\rightarrow$B$\rightarrow$Prop, subrelation R' R /\textbackslash{}
$\forall$ x : A, $\exists$! y : B, R' x y.
\end{description}

\section{Hall's Marriage Theorem\label{sec:Hall's-Marriage-Theorem}}

\subsection{Bipartite graphs}

A bipartite graph is a triple $(L,\,R\,,\,E)$ where $L\cap R=\phi$,
and $E$ consists of pairs from $L\times R$. Elements of $L\cup R$
are called vertices and elements of $E$ are called edges. Here, we
consider only finite bipartite graphs. In Coq, we define it as a dependent
record.
\begin{description}
\item [{Record}] Bipar\_Graph: Type := Def\_of\_BG \{ \\
Graph\_of\_BG:> Finite\_Graph ; \\
L\_of: Ensemble U; \\
R\_of: Ensemble U; \\
LR\_Inhabited: Inhabited \_ L\_of /\textbackslash{} Inhabited \_ R\_of;
\\
LR\_Disj: Disjoint \_ L\_of R\_of; \\
LR\_Union: Vertices\_of (Graph\_of\_BG ) = (Union \_ L\_of R\_of);
\\
LR\_Rel: $\forall$ x y: U, (Edge\_Rel\_of (Graph\_of\_BG )) x y $\rightarrow$
(In \_ L\_of x /\textbackslash{} In \_ R\_of y) \}.
\end{description}
Edges are defined as a binary relation on the vertices. 
\begin{itemize}
\item The neighborhood of a set $S\subset L$, denoted $N(S)$, is the set
of all those vertices that are in some edge containing a vertex from
$S$, i.e., \\
$N(S)=\{v\in R:\,\exists u\in L,\,(u,v)\in E\}$. 
\item A matching is a collection of disjoint edges, i.e., no two edges in
a matching have a common vertex. 
\item A matching is said to be $L$-perfect if each vertex in $L$ is part
of some edge of the matching. 
\end{itemize}
In Coq we define these terms as N (S), Is\_a\_matching and Is\_L\_Perfect.\texttt{ }The
exact definitions appear in \prettyref{sec:Appendix}(Appendix). 

\subsection{Hall's Marriage Theorem}

Let $G=(L,R,Edge)$ be a bipartite graph and $V=L\cup R$. Then we
have, 
\begin{description}
\item [{Theorem}] Halls\_Thm: ($\forall$(S: Ensemble U), Included \_ S
L $\rightarrow$ ($\forall$ m n :nat,(cardinal \_ S m /\textbackslash{}
cardinal \_ (N S) n) $\rightarrow$ m <=n ) ) $\leftrightarrow$ ($\exists$
Rel:Relation U, Included\_in\_Edge Rel /\textbackslash{} Is\_L\_Perfect
Rel).
\end{description}
where, Included\_in\_Edge is defined as,
\begin{description}
\item [{Definition}] Included\_in\_Edge (Rel: Relation U): Prop := $\forall$
x y:U, Rel x y $\rightarrow$ Edge x y.
\end{description}
It states that, for any bipartite graph $G=(L,R,E),\,\forall S\subset L,\,|N(S)|\geq|S|$
if and only if $\exists$ an $L$-perfect matching. \\
\textbf{Proof} : We prove the ``only if'' (forward direction) part
of the theorem, the ``if'' part being trivial. Once we have Dilworth's
theorem, a proof of Hall's theorem follows rather easily. Turn the
bipartite graph $(L,R,E)$ into a poset $\mathcal{P}$ whose elements
are vertices of $L\cup R$ and the relation is the reflexive closure
of the edge relation. One can imagine the bipartite graph as the Hasse
diagram of poset $\mathcal{P}$. 

First, we prove that $R$ is a largest antichain. Fix any antichain
$\mathcal{A=A_{L}\cup A_{R}}$ where $\mathcal{A_{L}},\mathcal{A_{R}}$
are in $L,R$ respectively. Now, $N(\mathcal{A_{L}})$ is disjoint
from $\mathcal{A_{R}}$ as $\mathcal{A}$ is an antichain. Hence,
$|\mathcal{A}|=|\mathcal{A_{L}}|+|\mathcal{A_{R}}|\leq|N(\mathcal{A_{L}})|+|\mathcal{A_{R}}|\leq|R|$.
Here the first inequality follows from the hypothesis $\forall S\subset L,\,|S|\leq|N(S)|$. 

Now, from Dilworth's theorem, there is a chain cover $\mathcal{C}$
of size $|R|$. Without loss of generality, the chains in $\mathcal{C}$
are disjoint. Each chain has to have an element of $R$. If we restrict
attention to the two element chains in $\mathcal{C}$, they form an
$L$-perfect matching. $\square$ 

Note that in the above proof Dilworth's theorem only assures the existence
of a chain cover $\mathcal{C}$ of size $|R|$. However, we claim
that without loss of generality the chains in $\mathcal{C}$ are disjoint.
This is a hidden assumption and needs a justification in the formal
proof. Just by looking at the informal proof of Hall's theorem one
might consider proving the following statement which justifies the
claim,
\begin{itemize}
\item In any finite poset P, if $\mathcal{C}$ is a chain cover of size
$|R|$ then there exists a disjoint chain cover $\mathcal{C}'$ of
size $|R|$ . 
\end{itemize}
It however turns out that the above statement is too strong. For example,
let $P=(C,R)$ be a poset where $C=\{a,b,c\}$ and $R$ is the reflexive
and transitive closure of the binary relation $R'=\{(a,b)\}$. Now
consider $\mathcal{C}=\{\{a\},\{b\},\{c\},\{a,b\}\}$, it is clearly
a chain cover of size 4. However, there can't be a disjoint chain
cover of size 4 for the poset P. Therefore, we consider the following
weaker statement, 
\begin{itemize}
\item In any finite poset P, if $\mathcal{C}$ is a smallest chain cover
of size $|R|$ then there exists a disjoint chain cover $\mathcal{C}'$
of size $|R|$.
\end{itemize}
This statement is proved as Lemma-17 in \prettyref{sec:Some-useful-results}.
Since Dilworth's theorem assures the existence of a smallest chain
cover $\mathcal{C}$ of size $|R|$ we use Lemma-17 in the formal
proof of Hall's Marriage theorem to justify the existence of a disjoint
chain cover. 

\subsection*{Sequence of distinct representative (SDR) }

The Hall's theorem on bipartite graph can be used to prove the original
form of Hall's theorem which talks about the representation of each
set in a collection of finite sets. Let $S=\{S_{1},\dots,S_{n}\}$
be a family of sets and $X=\underset{i\leq n}{\cup}S_{i}$. 
\begin{itemize}
\item A sequence of distinct representatives (SDR) for $S$ is a sequence
$\{x_{1},\dots,x_{n}\}$ of pairwise distinct elements of X such that
$x_{i}\in S_{i},1\leq i\leq n$. 
\end{itemize}
Hall's Marriage theorem then states that,
\begin{itemize}
\item $S$ has an SDR iff the union of any $k$ members of $S$ contains
at least $k$ elements. 
\end{itemize}
The above result easily follows from Hall's theorem on graphs. Consider
the bipartite graph $(S,X,E)$ where $E$ consists of all the pairs
$(S_{i},a_{i})$ where $a_{i}$ is a member of set $S_{i}$. An L-perfect
matching in this graph corresponds to an SDR for $S$ and the neighborhood
$N(S)$ becomes $\underset{S_{i}\in S}{\cup}S_{i}$. Hence the above
statement gets transformed to the statement of Hall's theorem on graphs. 

We closely follow this line of reasoning to prove the SDR version
of Hall's theorem in Coq. However, instead of considering a sequence
of distinct representatives we consider a relation that assures the
SDR criterions. It reduces the overheads of dealing with sequences.
In this setting we have,
\begin{description}
\item [{Theorem}] The\_Halls\_Thm: exists\_a\_one\_one\_map $\leftrightarrow$
union\_is\_at\_least\_m. 
\end{description}
where, 
\begin{description}
\item [{exists\_a\_one\_one\_map}] is an abbreviation for, ( $\exists$
Rel': Ensemble U $\rightarrow$ U$\rightarrow$ Prop, ($\forall$
(x:Ensemble U) (y:U), Rel' x y $\rightarrow$ In \_ x y) /\textbackslash{}
( $\forall$ (x y:Ensemble U) (z: U), (Rel' x z /\textbackslash{}
Rel' y z)$\rightarrow$ x=y) /\textbackslash{} ($\forall$ x: Ensemble
U, In \_ S x $\rightarrow$ ($\exists$ y: U, Rel' x y))) and,
\item [{union\_is\_at\_least\_m}] is an abbreviation for, ($\forall$ S':
Ensemble (Ensemble U), Included \_ S' S $\rightarrow$ ( $\forall$
m n:nat, (cardinal \_ S' m /\textbackslash{} cardinal \_ (Union\_over
S') n) $\rightarrow$ m<= n) ) 
\end{description}
Note that the existence of such a relation Rel'\texttt{ }assures the
existence of a one-one map from $S$ to $X$. Moreover, Rel' is contained
in the set membership relation; because Rel' x y $\rightarrow$ In
\_ x y. Hence, the existence of such relation implies the existence
of an SDR and vice-versa. 

There is one technical difficulty that arises while proving Hall's
theorem on sets using Hall's theorem on Bipartite graphs. In a Bipartite
graph the members of sets $L$ and $R$ are of the same type. It is
essential to define them in this way since constructing a poset from
graph and applying Dilworth's theorem becomes easy. However, in the
Hall's theorem on sets (SDR) the members of the sets $S$ and $X$
are of different types. The members of $S$ are of type $Ensemble\,U$
while the members of $X$ are of type $U$. 

Therefore it becomes difficult to prove The\_Halls\_Thm directly using
Halls\_Thm. To resolve this issue we consider a bipartite graph where
the left and right vertices are of different types. Let, 
\begin{description}
\item [{Variable}] L: Ensemble U.
\item [{Variable}] R: Ensemble V.
\item [{Variable}] Rel: U$\rightarrow$ V$\rightarrow$ Prop.
\end{description}
In this context we then prove the following statement, 
\begin{description}
\item [{Theorem}] Marriage\_Thm: ($\forall$ (S:Ensemble U), Included \_
S L $\rightarrow$ ($\forall$ m n :nat, (cardinal \_ S m /\textbackslash{}
cardinal \_ (Ngb S) n) $\rightarrow$ m <=n ) ) $\leftrightarrow$
( $\exists$ Rel': U$\rightarrow$ V$\rightarrow$ Prop, Included\_in\_Rel
Rel' /\textbackslash{} Is\_L\_Perfect\_matching Rel').
\end{description}
where Ngb and Is\_L\_Perfect\_matching are defined as, 
\begin{description}
\item [{Definition}] Ngb (S: Ensemble U):= fun (y: V) $\Rightarrow$ $\exists$
x:U, In \_ S x /\textbackslash{} Rel x y. 
\item [{Definition}] Is\_L\_Perfect\_matching (Rel: U$\rightarrow$ V$\rightarrow$
Prop):= ($\forall$ x: U, In \_ L x $\rightarrow$ ($\exists$ y:
V, In \_ R y /\textbackslash{} Rel x y)) /\textbackslash{} ( $\forall$
(x y: U)(z: V), (Rel x z /\textbackslash{} Rel y z) $\rightarrow$
x=y).
\end{description}
We prove this statement in the file Marriage\_Thm.v. Once we have
the above result it can be directly used to prove The\_Halls\_Thm
on sets. Proof of The\_Halls\_Thm using Marriage\_Thm also appears
in the same file Marriage\_Thm.v. 

\section{Sequences and the Erd\H{o}s-Szekeres Theorem\label{sec:Sequences-and-the}}

\subsection{Finite Sequence of Integers}

A sequence $(C,\prec)$ consists of a non-empty set $C$ together
with a binary relation $\prec$ satisfying asymmetry and transitivity
properties. Moreover, any two distinct elements of $C$ must be related
with this ordering relation. Note the difference with partial orders,
the relation $\prec$ is asymmetric instead of being antisymmetric.
This means for any two elements $a,\,b\in C$, $a\prec b\rightarrow\sim b\prec a$.
We define a sequence of integers in Coq as a dependent record,
\begin{description}
\item [{Record}] Int\_seq:Type:= Def\_of\_seq \{\\
C\_of: Ensemble nat; \\
R\_of: Relation nat;\\
Seq\_cond1: Inhabited \_ (C\_of); \\
Seq\_cond2: Finite \_ (C\_of); \\
Seq\_cond3: Transitive \_ R\_of; \\
Seq\_cond4: Asymmetric \_ R\_of; \\
Seq\_cond5: Total\_Order R\_of C\_of ; \}.
\end{description}
Since we are working only with finite sequences we declare it as Seq\_cond2
in the definition of Int\_seq. 

\subsection{The Erd\H{o}s-Szekeres Theorem}

For a finite sequence s: Int\_seq we prove, 
\begin{description}
\item [{Theorem}] Erdos\_Szeker: $\forall$ m n, cardinal (C\_of s) (m{*}n+1)
$\rightarrow$ (($\exists$ s1: Int\_seq, sub\_seq s1 s /\textbackslash{}
Is\_increasing s1 /\textbackslash{} cardinal (C\_of s1) (m+1)) \textbackslash{}/
($\exists$ s2: Int\_seq, sub\_seq s2 s /\textbackslash{} Is\_decreasing
s2 /\textbackslash{} cardinal (C\_of s2) (n+1))).
\end{description}
Here Is\_increasing and Is\_decreasing capture the notions of increasing
and decreasing sequences respectively. That s1 is a subsequence of
s2 is represented by predicate sub\_seq s1 s2.\texttt{ }The exact
definitions of these terms are given in \prettyref{sec:Appendix}(Appendix). 

The Erd\H{o}s-Szekeres theorem then states that for any two natural
numbers $m$ and $n$, every sequence of $m.n+1$ distinct integers
contains an increasing subsequence of length $m+1$ or a decreasing
subsequence of length $n+1$. \\
\textbf{Proof}: Let $(C,\prec)$ be the sequence where $|C|=m.n+1$.
To prove this theorem, we construct a poset $(C,\le)$ where for any
two $x,y\in C$, $x\le y$ iff $x\prec y$ and $x$ is less than $y$
as numbers. Note that, 
\begin{itemize}
\item A chain in this partial order $(C,\le)$ is a monotonically increasing
subsequence in $(C,\prec)$, and 
\item An antichain in $(C,\le)$ is a monotonically decreasing subsequence
in $(C,\prec)$ .
\end{itemize}
Now, we complete the proof of Erd\H{o}s-Szekeres theorem by proving
the following result on general posets, 
\begin{itemize}
\item If P is a poset with $m.n+1$ elements, then it has a chain of size
at least $m+1$ or an antichain of size at least $n+1$. 
\end{itemize}
This statement is proved as Lemma-20 in \prettyref{sec:Some-useful-results}.
It follows easily from the Dilworth's theorem. There can be two cases;
either there is an antichain $\mathcal{A}$ of size $n+1$ or the
size of a largest antichain is $n$. In the first case statement is
trivially true. In the second case, using Dilworth's theorem we know
that there exists a chain cover $\mathcal{C_{V}}$ of size $n$. Since
$\mathcal{C_{V}}$ covers the whole poset P and its size is $m.n+1$,
there must be a chain of size at least $m+1$ in $\mathcal{C_{V}}$.
This completes the proof. $\square$

\section*{Wrapping Up}

This work is done in the Coq Proof General (Version 4.4pre). We have
used the Company-Coq extension \cite{key-15} for the Proof General.
The proofs are split into following files:
\begin{enumerate}
\item \texttt{PigeonHole.v}: It contains some variants of the Pigeonhole
Principle.
\item \texttt{BasicFacts.v}: Contains some useful properties on numbers
and sets. It also contains strong induction and some variants of Choice
theorem. 
\item \texttt{FPO\_Facts.v}: Most of the definitions and some results on
finite partial orders are proved in this file. 
\item \texttt{FPO\_Facts2.v}: Contains most of the lemmas that we discussed
in this section. 
\item \texttt{FiniteDilworth\_AB.v}: Contains the proofs of forward and
backward directions of Dilworth's theorem.
\item \texttt{FiniteDilworth.v}: Contains the proof of the main statement
of Dilworth's theorem. 
\item \texttt{Combi\_1.v}: Some new tactics are defined to automate the
proofs of some trivial facts on numbers, logic, sets and finite partial
orders. 
\item \texttt{BasicFacts2.v: }Contains some facts about power-sets.
\item \texttt{FPO\_Facts3.v:} Contains some more lemmas on finite posets.\texttt{ }
\item \texttt{Dual\_Dilworth.v}: Contains the proof of the Dual-Dilworth
Theorem. 
\item \texttt{Graph.v:} Contains definitions of different types of graphs.
\item \texttt{Halls\_Thm.v: }Contains the proof of Hall's theorem on bipartite
graph. 
\item \texttt{Marriage\_Thm.v:} Contains the proof of Hall's theorem on
collection of finite sets (SDR).
\item \texttt{Erdos\_Szeker.v:} Contains the proof of the Erd\H{o}s-Szekeres
theorem on sequences.
\end{enumerate}
The Coq code for this work is available at \cite{key-16}. The files
can be safely compiled in the given order. 

\section{Related Work\label{sec:Related-Work}}

Rudnicki \cite{key-11} presents a formalization of Dilworth's decomposition
theorem in Mizar. In the same paper they also provide a proof of the
Erd\H{o}s-Szekeres theorem using Dilworth's theorem. A separate proof
of the Hall's marriage theorem in Mizar appeared in \cite{key-12}.
Jiang and Nipkow \cite{key-13} also presented two different proofs
of Hall's theorem in Isabelle/HOL. We have used a different theorem
prover and formalized all of these results in a single framework.
Our work is closest to the work of \cite{key-11}. However, we added
extra results (Hall's theorem) in the same framework. The proof we
mechanize for Hall's theorem uses Dilworth's theorem and we formalize
Hall's theorem in both of its popular forms. The first form deals
with the matching in a bipartite graph and the second form is about
sequence of distinct representatives (SDR) for a collection of finite
sets. We also provide a clear compilation of some useful results on
finite sets and posets that can be used for mechanizing other important
results from the combinatorics of finite structures. 

\section{Conclusions\label{sec:Conclusions}}

Formalization of any mathematical theory involves significant time
and effort because the size of formal proofs blows up significantly.
In such circumstances exploring dependencies among important results
might save some time and effort. Dilworth's decomposition theorem
is an important result on partially ordered sets (poset). It has been
used successfully to give concise proofs of some other important results
from combinatorics. Here we use Dilworth's theorem on posets to mechanize
proofs of two other well known results on sets and sequences. The
main contributions of this paper are: 
\begin{enumerate}
\item Fully formalized proofs of Dilworth's decomposition theorem and Mirsky's
theorem in Coq, together with an explanation of all the definitions
and the theorem statement. 
\item Fully mechanized proofs of Hall's Marriage theorem and the Erd\H{o}s-Szekeres
theorem using Dilworth's decomposition theorem. 
\item A clear compilation of some general results and definitions which
could be used as a framework in the formalization of other similar
results. 
\end{enumerate}
The Coq code for this work is available at \cite{key-16}. One can
further explore the dependencies of these mechanized results with
other well known results in combinatorics. It can save a lot of time
and effort in mechanizing their proofs.

\appendix

\section{Appendix\label{sec:Appendix}}

\subsection*{Partial Orders, chains and antichains}

For a finite partial order P: FPO U on some type \texttt{U} let, 

C := Carrier\_of U P and,

R:= Rel\_of U P.\texttt{ }

Then, we have the following definitions: 
\begin{enumerate}
\item Definition \emph{Is\_a\_chain\_in} (e: Ensemble U): Prop:= (Included
U e C /\textbackslash{} Inhabited U e) /\textbackslash{} ($\forall$
x y:U, (Included U (Couple U x y) e) $\rightarrow$ R x y \textbackslash{}/
R y x).\texttt{ }
\item Definition \emph{Is\_an\_antichain\_in} (e: Ensemble U): Prop := (Included
U e C /\textbackslash{} Inhabited U e) /\textbackslash{} ($\forall$
x y:U, (Included U (Couple U x y) e) $\rightarrow$ (R x y \textbackslash{}/
R y x) $\rightarrow$ x=y).
\item Inductive \emph{Is\_largest\_chain\_in} (e: Ensemble U): Prop:= largest\_chain\_cond:
Is\_a\_chain\_in e $\rightarrow$ ($\forall$ (e1: Ensemble U) (n
n1:nat), Is\_a\_chain\_in e1 $\rightarrow$ cardinal \_ e n $\rightarrow$
cardinal \_ e1 n1 $\rightarrow$ n1 $\leq$ n) $\rightarrow$ Is\_largest\_chain\_in
e.
\item Inductive \emph{Is\_largest\_antichain\_in} (e: Ensemble U): Prop:=
largest\_antichain\_cond: Is\_an\_antichain\_in e $\rightarrow$ ($\forall$
(e1: Ensemble U) (n n1: nat), Is\_an\_antichain\_in e1 $\rightarrow$
cardinal \_ e n $\rightarrow$ cardinal \_ e1 n1 $\rightarrow$ n1
$\leq$ n ) $\rightarrow$ Is\_largest\_antichain\_in e. 
\item Inductive \emph{Is\_a\_chain\_cover}(cover:Ensemble(Ensemble U)):
Prop:= cover\_cond: ($\forall$ (e: Ensemble U), In \_ cover e $\rightarrow$
Is\_a\_chain\_in e) $\rightarrow$ ($\forall$ x:U, In \_ C x $\rightarrow$
($\exists$ e: Ensemble U, In \_ cover e /\textbackslash{} In \_ e
x)) $\rightarrow$ Is\_a\_chain\_cover cover.
\item Inductive \emph{Is\_an\_antichain\_cover} (cover: Ensemble (Ensemble
U)): Prop:= AC\_cover\_cond: ($\forall$ (e: Ensemble U), In \_ cover
e $\rightarrow$ Is\_an\_antichain\_in e) $\rightarrow$ ($\forall$
x:U, In \_ C x $\rightarrow$ ($\exists$ e: Ensemble U, In \_ cover
e /\textbackslash{} In \_ e x)) $\rightarrow$ Is\_an\_antichain\_cover
cover.
\item Inductive \emph{Is\_a\_smallest\_chain\_cover} (scover: Ensemble (Ensemble
U)): Prop:= smallest\_cover\_cond: (Is\_a\_chain\_cover P scover)
$\rightarrow$ ($\forall$(cover: Ensemble (Ensemble U)) (sn n: nat),
(Is\_a\_chain\_cover P cover /\textbackslash{} cardinal \_ scover
sn /\textbackslash{} cardinal \_ cover n) $\rightarrow$ (sn $\leq$
n)) $\rightarrow$ Is\_a\_smallest\_chain\_cover P scover.
\item Inductive \emph{Is\_a\_smallest\_antichain\_cover} (scover: Ensemble
(Ensemble U)): Prop:= smallest\_cover\_cond\_AC: (Is\_an\_antichain\_cover
P scover) $\rightarrow$ ($\forall$(cover: Ensemble (Ensemble U))
(sn n: nat), (Is\_an\_antichain\_cover P cover /\textbackslash{}cardinal
\_ scover sn /\textbackslash{} cardinal \_ cover n) $\rightarrow$
(sn $\leq$ n)) $\rightarrow$ Is\_a\_smallest\_antichain\_cover P
scover.
\item Inductive \emph{Is\_height }(n: nat) : Prop:= H\_cond: ($\exists$
lc: Ensemble U, Is\_largest\_chain\_in P lc /\textbackslash{} cardinal
\_ lc n) $\rightarrow$ (Is\_height P n). 
\item Inductive \emph{Is\_width} (n: nat) :Prop := W\_cond: ($\exists$
la: Ensemble U, Is\_largest\_antichain\_in P la /\textbackslash{}
cardinal \_ la n) $\rightarrow$ (Is\_width P n).
\end{enumerate}

\subsection*{Bipartite graphs and matching}
\begin{enumerate}
\item Definition \emph{N }(S: Ensemble U): Ensemble U:= fun (y: U) $\Rightarrow$
$\exists$ x:U, In \_ S x /\textbackslash{} Edge x y.
\item Definition \emph{Is\_a\_matching} (R: Relation U): Prop:= ( $\forall$
x y z: U, ((R x z /\textbackslash{} R y z)\textbackslash{}/ (R z x
/\textbackslash{} R z y)) $\rightarrow$ x=y). 
\item Definition \emph{Is\_L\_Perfect} (Rel: Relation U): Prop:= (Is\_a\_matching
Rel /\textbackslash{} ($\forall$ x: U, In \_ L x $\rightarrow$ ($\exists$
y: U, Rel x y))).
\end{enumerate}

\subsection*{Increasing and decreasing subsequences}
\begin{enumerate}
\item Definition \emph{Asymmetric} := fun (U : Type) (R : Relation U) $\Rightarrow$
$\forall$ x y : U, R x y $\rightarrow$ \textasciitilde{} R y x. 
\item Definition \emph{Total\_Order} (U:Type )(R: Relation U)(S: Ensemble
U): Prop:= $\forall$ s1 s2, (In \_ S s1 /\textbackslash{} In \_ S
s2) $\rightarrow$ ( R s1 s2 \textbackslash{}/ R s2 s1).
\item Definition \emph{sub\_seq} (s1 s2: Int\_seq): Prop:= Included \_ (C\_of
s1) (C\_of s2)/\textbackslash{} ($\forall$ m n, (In \_ (C\_of s1)
m /\textbackslash{} In \_ (C\_of s1) n ) $\rightarrow$ R\_of s1 m
n $\rightarrow$ R\_of s2 m n ).
\item Definition \emph{Is\_increasing} (s: Int\_seq): Prop:= $\forall$
m n, (In \_ (C\_of s) m /\textbackslash{} In \_ (C\_of s) n ) $\rightarrow$
R\_of s m n $\rightarrow$ m < n.
\item Definition \emph{Is\_decreasing} (s: Int\_seq): Prop:= $\forall$
m n, (In \_ (C\_of s) m /\textbackslash{} In \_ (C\_of s) n ) $\rightarrow$
R\_of s m n $\rightarrow$ m > n. 
\end{enumerate}


\begin{thebibliography}{10}
\bibitem{key-1} R. P. Dilworth. \newblock A Decomposition Theorem
for Partially Ordered Sets, \newblock\emph{ Annals of Mathematics,
}Vol. 51 (1951), pp. 161-66.

\bibitem{key-2}M. A. Perles, \newblock A Proof of Dilworth's Decomposition
Theorem for Partially Ordered Sets, \newblock\emph{ Israel J. Math,
}1963, pp. 105- 107.

\bibitem{key-3} H. Tverberg, \newblock On Dilworth\textquoteright s
decomposition theorem for partially ordered sets, \newblock\emph{
J. Combin. theory, }1967, pp. 305-306.

\bibitem{key-4}F. Galvin, \newblock A proof of Dilworth's chain
decomposition theorem, \newblock\emph{ American Mathematical Monthly,
}Vol. 101, No. 4 (1994), pp. 352\textendash 353.

\bibitem{key-5} Philip Hall, \newblock On Representations of Subsets,
\newblock\emph{ J. London Math. Soc. }10(1), pp. 28-30.

\bibitem{key-6} Paul R. Halmos and Herber E Vaughan, \newblock The
marriage problem, \newblock\emph{ American Journal of Mathematics
72}, pp. 214-215, 1950.

\bibitem{key-7} Martin Aigner and Gnter M. Ziegler. \newblock Proofs
from THE BOOK (4th ed.). \newblock\emph{ Springer Publishing Company,}
Incorporated. 

\bibitem{key-8} P. Erd\H{o}s and G. Szekeres, \newblock A combinatorial
problem in geometry. \newblock\emph{ Compositio Mathematica,} 2:463\textendash 470,
1935. 

\bibitem{key-9} The Coq development team, \newblock The Coq proof
assistant reference manual, \newblock 2016, v8.5. 

\bibitem{key-10} Y. Bertot and P. Casteran, \newblock Interactive
Theorem Proving and Program Development. \newblock Coq'Art: The Calculus
of Inductive Constructions Series: Texts in Theoretical Computer Science.
2004.

\bibitem{key-11} P. Rudnicki, \newblock Dilworth's Decomposition
Theorem for Posets, \newblock\emph{ Formalized Mathematics Vol.17,
No. 4, }pp. 223-232, 2009. 

\bibitem{key-12} E. Romanowicz, Adam Grabowski, \newblock The Hall
Marriage Theorem, \newblock\emph{ Formalized Mathematics Vol.12,
No.3, }pp. 315-320, 2004.

\bibitem{key-13} D. Jiang, Tobias Nipkow, \newblock Proof Pearl:
The Marriage Theorem, \newblock\emph{ Certified Programs and Proofs:
First International Conference, CPP 2011, Kenting, Taiwan, Dec 7-9,
2011. }Proceedings. pp. 394-399, 2011.

\bibitem{key-14} The Coq Standard Library, https://coq.inria.fr/distrib/current/stdlib/.

\bibitem{key-15}Clément Pit-Claudel and Pierre Courtieu, Company-Coq:
Taking Proof General one step closer to a real IDE, \emph{CoqPL'16:
The Second International Workshop on Coq for PL, }2016.

\bibitem{key-16} Diworth, Hall and Erd\H{o}s-Szekeres in Coq, http://www.tcs.tifr.res.in/\textasciitilde{}abhishek/.

\bibitem{key-17}Leon Mirsky, A dual of Dilworth's decomposition theorem,
\emph{American Mathematical Monthly,} Vol. 78, No. 8 (1971), pp. 876\textendash 877.

\bibitem{key-18}H. Geuvers, Proof assistants: History, ideas and
future, \emph{Sadhana, Vol. 34, No. 1}, pp. 3-25, 2009. 

\bibitem{key-19}Wikipedia: K\H{o}nig's theorem \textemdash{} https://en.wikipedia.org/wiki/K\H{o}nig's\_theorem\_(graph\_theory).
\end{thebibliography}
\end{document}